\documentclass{article}
\usepackage[utf8]{inputenc}
\usepackage[title]{appendix}
\usepackage{bbm}
\usepackage{amssymb}
\usepackage[english]{babel}    
\usepackage{authblk}
\usepackage{amsmath,amsthm}           
\usepackage[utf8]{inputenc}    
\usepackage[T1]{fontenc}       
\usepackage{longtable}         
\usepackage{exscale}           
\usepackage[final]{graphicx}   
\usepackage[sort]{cite}        
\usepackage{array}             
\usepackage{eucal} 
\usepackage{hyperref}
\usepackage{wasysym}           
\usepackage[a4paper]{geometry} 
\usepackage[multiuser]{fixme}  
\usepackage{xspace}            
\usepackage{tikz}              
\usepackage{ifdraft}           
\usepackage[expansion=false    
           ]{microtype}        
\usepackage[nottoc]{tocbibind} 
\usepackage[all]{xy}
\usepackage{slashed}
\usepackage{cleveref}
\usetikzlibrary{matrix}
\usepackage{tikz-cd}
\usepackage{amsthm}

\newtheorem{remark}{Remark}[section]

\title{Beyond Signal and Noise: Unraveling Scale Invariance in Neuroscience and Financial Networks with Topological Data Analysis}

\author[1]{Roel Gisolf\thanks{email: roel.gisolf@gmail.com}}
\author[1,2,3]{Fernando A. N. Santos\thanks{email: f.a.nobregasantos@uva.nl}}
\author[4]{Felix Wierstra\thanks{email: felix.wierstra@gmail.com}}
\affil[1]{Korteweg-de Vries Institute for Mathematics, University of Amsterdam, Science Park 105-107, Amsterdam, The Netherlands}
\affil[2]{Dutch Institute for Emergent Phenomena (DIEP), 1090 GL Amsterdam, The Netherlands}
\affil[3]{Amsterdam UMC location Vrije Universiteit Amsterdam, Anatomy and Neurosciences, Boelelaan 1117, Amsterdam, The Netherlands}
\affil[4]{Mathematisch Instituut, Universiteit Utrecht, Budapestlaan 6, The Netherlands}
\date{}
\begin{document}
\maketitle

\begin{abstract}
Topological Data Analysis (TDA) is increasingly crucial in investigating the shape of complex data structures across scientific fields, particularly in neuroscience and finance. This study delves into persistent homology, a TDA component initially aimed at differentiating between signal and noise. We explore two methodologies: the conventional cycle length approach and the novel death-birth ratio method proposed by Bobrowski and Skraba.
Analyzing rs-fMRI data from the Human Connectome Project and daily $S\&P 500$ financial networks, our study compares these methods in identifying significant cycles. A key discovery is a robust relationship between z-score thresholds applied to bar lengths or ratios and behavioural traits in brain networks and market volatility in financial networks. In the brain, this is evident in the strong correlation between significant 1-cycles, brain volumes, and sex-based differences. In financial markets, a fractal pattern emerges, where market volatility negatively correlates with the number of significant cycles, indicating that more complex market topologies are associated with increased stability.
Our findings also imply a fractal nature of 1-cycles at both population levels and across multiple days in the stock market. The distribution of significant loops, marked by high z-scores, remains consistent across various z-score thresholds, revealing a scale-invariant, fractal structure in both data sets. Given the scale invariance in these fractal structures, the traditional TDA distinction between signal and noise becomes less meaningful. This suggests that all scales of cycle length are relevant, challenging the conventional approach of segregating signal from noise and broadening the scope of TDA to reveal intricate, scale-invariant relationships in complex systems.

\end{abstract}

\section{Introduction}



Topological Data Analysis (TDA) has emerged over the last decade as a powerful tool in data analysis, employing topological techniques to explore the complex shapes and geometric properties of data \cite{Zomorodian,carlsson2009topology,Otteretal}. Its robustness to noise \cite{Cohenetal} makes TDA a reliable approach in a variety of applications. This attribute has led to its increasing use in areas such as neuroscience \cite{santos2019topological}, financial modelling \cite{majumdar2020clustering}, network analysis and complex systems \cite{patania2017topological,de2022euler}. TDA’s ability to reveal underlying structures in data that might otherwise remain hidden has contributed to its growing prominence in empirical research. The expanding literature reflects its practical utility and the evolving interest in its applications. As a method, TDA offers a unique perspective in understanding the intricate shape of complex datasets, marking a significant advancement in how we analyze and interpret data \cite{carlsson2009topology}.


Persistent homology is a key technique in TDA, primarily used for analyzing the shape and structure of data, typically represented as point clouds \cite{edelsbrunner2002topological}. This method constructs a nested sequence of simplicial complexes, like the Vietoris-Rips complex, to systematically examine the homology of data. The core idea is to count the number of 'holes' or cycles, providing insight into the topological and geometric properties of the dataset.

In mathematical terms, this involves investigating the homology of these shapes. The outcome of this process is represented in a persistence diagram or 'bar code', which displays homology classes along with their emergence (birth) and disappearance (death) times \cite{ghrist2008barcodes}. This approach helps in distinguishing between persistent features that likely represent true structures (signal) in the data and transient ones that might be attributed to noise. However, interpreting these diagrams requires careful consideration on what is signal and what is noise to avoid overestimating the importance of certain topological features across scales.

In the process of creating such a persistence diagram, it is often assumed that among the homology classes created, some correspond to features of the data set and a large number are noise. Therefore, one of the main challenges in persistent homology is to determine which homology classes correspond to actual information (signal) about the data set and which ones are noise. The most common method to do this is by looking at the lifetime ($\Delta=\mbox{time of death} - \mbox{time of birth}$) of each class and assuming that the longer-lived ones correspond to actual homology classes and the short-lived ones correspond to noise. In many cases, one could say that just looking at the lifespan of the classes gives good estimates of signal and noise in point clouds. 


Another solution, proposed by Bobrowski, Kahle and Skraba (see \cite{BobrowskiKahleSkraba}), is to look at the death-birth ratio $\pi=(\mbox{time of death}/\mbox{time of birth})$, instead of the difference. The key idea here is that this is scale-invariant (see Figure 4 of \cite{BobrowskiSkraba} for more details) and favours dense subsets in the point cloud compared to sparse ones (see Figure 6 of \cite{BobrowskiSkraba}). 
Bobrowski and Skraba further conjectured that, after a transformation, the distribution of noise ratios follows a specific distribution called the LGumbel or left-skewed Gumbel distribution. If this conjecture turns out to be true, then every homology class in the persistence diagram that significantly differs from the LGumbel distribution  would correspond to an actual homology class, possibly distinct from noise, and therefore a feature of the data.

The initial goal of this paper was to test the ideas of Bobrowski and Skraba \cite{BobrowskiSkraba} on two data sets - in neuroscience and finance respectively - and give a comparison between the two methods of distinguishing cycles. However, we realized that this exploratory analysis lead us to new insights on scale invariant properties of persistent diagrams, as we will show in this work.

The first data set comes from the Human Connectome Project (HCP) and consists of resting state fMRI brain scans of $1084$ healthy individuals. We compute the persistent homology per person and apply the two different statistical signal-noise interpretations. We further compare the number of $1$-cycles of these tests with certain annotated behavioural traits, more specifically, brain volumes and sex. We find that there is a correlation of $-0.303$ between the total number $1$-cycles and the brain volume in the HCP data. It further turns out that on average the number of loops in the female part of the population is higher than in the male part (see Figure \ref{fig:brainvolumecorrelation}).

The second data set is the time series of the stock prices of the S\&P 500. Again, we compute the  persistent homology and compare the results of the different statistical tests. We further compute the correlation between the volatility of the S\&P 500 and our results for the number of significant 1-homologies. Here, we find that there is a correlation of $-0.71$ between the total number of $1$-cycles in the financial network associated to the S\&P 500 and the daily volatility (see Figure \ref{fig:volatilitycorrelation}). 

In both cases, we found that all statistical tests for distinguishing signal and noise give strongly correlated results and that there was very little difference between the persistence length and the persistence ratios in real data. We further found that taking a cutoff for distinguishing signal and noise at different $z$-scores  gave very similar results. This scale invariance leads us to believe that the distribution of lengths and ratios of these data sets have fractal properties, which would explain why all the tests were so similar. We therefore computed the Hurst exponent and applied the box counting method to each dataset. This computation gave us strong evidence that the distribution of ratios both data sets are indeed fractal. 

Remarkably, this seems to suggest that in real, complex data, the interpretation of persistent homology is independent of statistical tests and $z$-scores chosen, and that there is thus no clear distinction between what is signal and what is noise. To summarize, this paper shows that both the HCP and S\&P 500 data sets have fractal properties, which we suspect qualitatively explains why all statistical tests give similar results. 

Since most experimental data has fractal properties, we suspect that by studying the persistent homology of fractals and fractal data has the potential to be used simplify the use of TDA in experimental data. 
In future work, we will explore this connection further in theoretical models.

The structure of this paper is as follows: Section 2 provides an essential background on Topological Data Analysis, laying the foundation for our methodologies. In Section 3, we delve into the analysis of functional brain networks, utilizing data from the Human Connectome Project. Section 4 shifts the focus to the financial networks, examining networks derived from S\&P 500 data in the year 2023. We conclude in Section 5 with a summary of our findings and discussions on the implications of our research. 

\section{Background and methods}

In this section, we recall the necessary preliminaries on persistent homology, and fractals and power laws. First, we recall the definition of persistent homology and explain several methods of distinguishing signal from noise. Then we recall several methods to test for fractality. These results will be necessary in Sections \ref{sec:HCP} and \ref{sec:financialdata} in which we show that the data from the HCP and S\&P 500 are fractal on a population level.

\subsection{Background on topological data analysis}

We briefly recall some notions from topological data analysis and explain how they are used in our experiments.  We assume that the reader is familiar with the basic definitions of simplicial complexes and persistent homology and only explain our conventions. For more details see for example  \cite{carlsson2009topology},  \cite{Otteretal},\cite{edelsbrunner2022computational}, and \cite{Zomorodian}.

\subsubsection{Persistent homology}

Persistent homology is a process that associates a sequence of homology groups to a point cloud, where a point cloud $X$ is a finite subset of points in a metric space (in our case, always $\mathbb{R}^n$ with the Euclidean metric). From a point cloud, we can construct a sequence of geometric objects called simplicial complexes, where a simplicial complex can be seen as a higher-dimensional version of a graph. The objects in this sequence are constructed using the Vietoris-Rips complex and are parameterized by a parameter $r > 0$. Roughly speaking, the Vietoris-Rips complex at $r$ is defined as the simplicial complex, which has a vertex set given by $X$ and has a $k$-simplex for every subset of $(k+1)$ points such that these $(k+1)$ points are at most distance $r$ from each other. (In this paper we have chosen to work with the Vietoris-Rips complex, but one could also use the Cech complex to get equivalent results.)

Persistent homology is then defined by taking the homology of this sequence of complexes, keeping track for which $r$ cycles are born and when they die \cite{edelsbrunner2002topological}. More precisely, the time of birth of a cycle $\tau$ is the smallest $r$ for which $\tau$ exists; similarly, the time of death is the largest $r$ for which $\tau$ is non-trivial. The results of taking persistent homology can be conveniently organized in what is called a persistence diagram. This is a multiset indexed by all the possible homology classes that exist at some point in time and consists of pairs $(\text{time of death}, \text{time of birth})$. For technical reasons, one usually also includes the diagonal, but this will not be relevant for this paper. The $n$th Betti number at time $r$ is defined as the rank of the $n$th homology group for that time.

The idea behind persistent homology is that by varying the parameter $r$, we run through all possible homologies or "shapes" the point cloud potentially could have. One of the main challenges of persistent homology is to determine which of these possible homologies are actual features of the data set, i.e., the underlying manifold associated with the data, and which ones are noise. To do this, one usually divide the persistence diagram $dgm$ into two parts, $dgm^S$, which corresponds to the signal, and $dgm^N$, which corresponds to the noise. The intuitive idea is that loops with a longer lifetime correspond to the actual signal, while loops with a short lifespan would correspond to noise. In practice, this means that one need statistical tests to decide which cycles are signal and which ones are noise. 

One of the main features of persistent homology is its robustness to noise, i.e. small perturbations in the point cloud only induce small differences in the persistence diagram. Under this perspective, small errors in the measurement, therefore do not change whether a cycle is signal or noise. The robustness of persistent homology was made precise in the Stability Theorem of Cohen-Steiner, Edelsbrunner and Harer (see \cite{Cohenetal}). 

In the remainder of this section, we describe the statistical tests that we compare in the remainder of the paper.

\subsubsection{The statistical interpretation of persistent homology}\label{sec:statisticalinterpretation}




To decide whether a cycle is significant and exclude it from being noise, we consider three statistical criteria. The first one is based on the lifespan of a cycle, while the second and third ones are based on the ratio between the time of death and the time of birth. The second one compares the death-birth ratio naively to the other death-birth ratios, while the third test is based on  recent conjectures by Bobrowski and Skraba \cite{BobrowskiSkraba}. In Sections \ref{sec:HCP} and \ref{sec:financialdata}, we apply all three of these methods to data from the Human Connectome Project and the timeseries of the S\&P 500, to give a comparison of their effectiveness in practice.

Since we want to compare different persistence diagrams coming from different people, we need to normalize our data. We do this by taking the $z$-scores of the length and ratios of the persistence intervals.

In all three tests, we start with a normalized persistence diagram with $n$ $k$-cycles, which is given by $dgm_k=\{ p_1,...,p_n\}$.

\subsubsection{Persistent length}

The first statistical method is by looking at the length of the persistent cycles. The length $\Delta$ of a cycle is defined by 
\[
\Delta=(\mbox{time of death}-\mbox{time of birth}).
\] 
So for every cycle $\tau=(\mbox{birth},\mbox{death})$ in our persistence diagram, we get a value $\Delta_\tau$. We normalize the set of persistence lengths by computing their $z$-scores. More precisely, we compute the mean $\mu$ and standard deviation $\sigma$ of the set of persistence lengths and normalize by sending $\Delta_\tau$ to $\frac{\Delta_\tau-\mu}{\sigma}$. The statistical test is then given by  picking a cutoff value for $z$ and excluding all the cycles whose $z$-scores are lower than that value. In practice, the value for $z_cutoff$  would be a critical $z$-score value where the distinction between signal and noise would be optimal. In practice, if the distinction between signal and noise holds, one might see a natural separation in the data, where a cluster of z-scores is distinctly higher than the rest, indicating potential signal. However, as we will see bellow, this is often not aways the case. 


\subsubsection{Persistence ratios}

A second statistical criteria to distinguish signal from noise is based on the distribution of death-birth ratio $\pi$, which is defined by 
\[
\pi=\frac{\mbox{time of death}}{\mbox{time of birth}}.
\]
This idea, which is due to Bobrowski, Kahle and Skraba (see \cite{BobrowskiKahleSkraba}) and was further developed in \cite{BobrowskiSkraba}, has several major theoretical advantages. The first theoretical advantage is that the ratio is scale invariant 
(see Figure 4 of \cite{BobrowskiSkraba}). The second advantage is that it is less sensitive to outliers. As is explained in Figure 6 of \cite{BobrowskiSkraba}, the persistence ratio favours subsets of the point cloud with a denser set of points. Since outliers are far away from the denser regions, they only create cycles with a relatively late birth. The resulting persistence ratio of a cycle involving an outlier would therefore be much smaller than that of cycles  not involving outliers. 

\begin{remark}
    Since every $0$-cycle is born at time $r=0$, this method does not work for the zeroth homology groups, since we would have to divide by zero. One might naively try to add a small perturbation $\epsilon$ to the birth of all $0$-cycles to avoid the division by zero. This idea does not work however since it would just be a linear transformation, which would give the same $z$-scores as the normalized persistent length.
\end{remark}

Using the persistence ratio, we can do two different statistical tests. The first "naive" option would be to first compute the $z$-scores of the persistence ratios, then pick a cutoff value and take the cycles whose $z$-scores are above this cutoff. Note that in this case, the $z$-scores are computed by taking the mean and standard deviation of the persistence ratios and not the lengths. This statistical test is then similar to the one using persistence lengths.

The second statistical test using the distribution of the persistence ratio is based on a conjecture of Bobrowski and Skraba. In \cite{BobrowskiSkraba}, they conjecture that the noise in a persistence diagram always follows a universal distribution given by the LGumbel or left-skewed Gumbel distribution. The CDF and PDF of this distribution are given by 
\[
F(x)=1-e^{-e^{x}} \mbox{ and } f(x)=e^{x-e^x}.
\]
Bobrowski and Skraba further describe a new statistical test in case their conjectures are true. This test claims that every cycle whose value is significantly different from the LGumbel distribution would be signal and the others would be noise.

In practice, this statistical test is defined as follows. Given the degree $k$ part of a persistence diagram $dgm_k=\{p_1,...,p_m\}$, we want to decide for every cycle $p_i$ whether it is noise or not. To do this, we first apply the transformation
\[
\ell(p)=\frac{1}{2} \log \log (\pi(p_i))-\lambda-\frac{1}{2}\frac{1}{\mid dgm_k\mid}\sum_{p \in dgm_k} \log \log (\pi(p)),
\]
where $\lambda $ denotes the Euler-Mascheroni constant ($= 0.5772156649...$) and $\mid dgm_k\mid$ the total number of $k$-cycles. The null hypothesis of the test is given by
\[
H^{(i)}_0:\ell(p_i) \sim \mbox{LGumbel}.
\]
If $x=\ell(p_i)$, is the observed persistence value then the corresponding $p$-value is computed as
\[
p\mbox{-value}_i= \mathbb{P}\left(\ell(p_i) \geq x \mid H^{(i)}_0 \right) = e^{-e^x}.
\]
If we want to test all cycles simultaneously, we need to apply the Bonferroni correction. For a fixed significance level $\alpha$, the signal part of the diagram is given by 
\[
dgm_k^S(\alpha)= \Bigl\{ p\in dgm_k: e^{-e^x} < \frac{\alpha}{\mid dgm_k\mid} \Bigl\}.
\]
This gives an alternative test that would possibly distinguish the noise from the signal in a persistence diagram. 

As we see later in the paper, applying the Bonferroni correction severely restricts the number of cycles passing the test. For small point clouds, essentially no cycles would pass the Bonferroni corrected test.

In the next few section we compare these three methods on data sets from the Human Connectome Project and timeseries form the S\&P 500.


\subsubsection{Z-Scores in Persistent Homology}

In statistics, and particulary in persistent homology, z-scores can be used for normalizing the persistence lengths or ratios of cycles, facilitating standardized comparisons across different datasets. This normalization is defined as follows:

\begin{equation}
z_\tau = \frac{\Delta_\tau - \mu}{\sigma},
\end{equation}

where \( \Delta_\tau \) represents the persistence length (or ratio) of the cycle, \( \mu \) is the mean persistence length (or ratio) of the dataset, and \( \sigma \) denotes the standard deviation. Z-scores enable us to determine whether a cycle is significantly longer-lived or has a significantly higher ratio compared to the average cycle, thereby indicating whether it is likely signal or noise. A threshold \( z_{\text{cutoff}} \) is then applied to classify cycles accordingly, indicating how many standard deviations a barcode lenght (or ratio) is compared with the average barcode.

\subsection{Fractals and power laws}

In our computations, we found that in both the HCP and S\&P 500 data there was certain  degree of fractality. We will now recall the two methods we use to determine whether the data was in fact fractal.

\subsubsection{Box Counting Method for Fractal Dimension}

The box counting method is utilized to estimate the fractal dimension of a dataset, which provides insights into the scale-invariant properties of the data. The method involves:

\begin{itemize}
    \item Dividing the space into a grid of boxes of side length \( \epsilon \).
    \item Counting the number of boxes \( N(\epsilon) \) that contain part of the data.
    \item Estimating the box-counting dimension \( D \) using the formula:
    \begin{equation}
    D = \lim_{\epsilon \to 0} \frac{\log N(\epsilon)}{\log(1/\epsilon)}.
    \end{equation}
\end{itemize}

This method is particularly relevant for assessing the fractal nature of persistence diagrams and understanding the complexity of the underlying topological features. For more details see for example Section 2 of \cite{PilgrimTaylor}.

\subsubsection{Hurst Exponent for Long-Term Memory Analysis}

The Hurst exponent \( H \) is a critical measure used to evaluate the presence of long-term memory and fractal-like behavior in time series data. For details on the definition of the Hurst exponent we refer the reader to for example  Section 4 of \cite{PilgrimTaylor}. The interpretation of \( H \) values is as follows:

\begin{itemize}
    \item \( H = 0.5 \): Indicates a random walk, typical of Brownian motion, with no long-term correlation.
    \item \( H > 0.5 \): Suggests long-term positive autocorrelation, implying that an increasing trend is likely to continue.
    \item \( H < 0.5 \): Indicates long-term negative autocorrelation, suggesting that an increasing trend is likely to reverse.
\end{itemize}

In this study, the Hurst exponent is used to determine whether the distribution of features in the persistence diagrams exhibits fractal characteristics, thereby suggesting scale invariance within the data.

These methodologies form the foundation for our analysis, enabling us to explore the scale-invariant properties and fractal structures in the datasets examined.

\section{Scale Invariance in Loops of Functional Brain Networks: Unveiling Behavioral and Sex-Based Correlations}\label{sec:HCP}

 Topological data analysis, particularly persistent homology, has been extensively applied in neuroscience across a wide range of scenarios, from analyzing brain connectivity patterns and understanding neural network dynamics to exploring the structural organization of the brain and identifying biomarkers for neurological disorders \cite{petri2014homological,giusti2016two,giusti2015clique,santos2019topological,bendich2016persistent,curto2017can} .In this work, we specifically employed the Schaefer Atlas with 1000 nodes plus 16 subcortical areas \cite{schaefer2018local}, as sourced from the Young Adult database of the Human Connectome Project's resting-state functional MRI scans, and freely available on Zenodo \cite{ZENODO}. This choice was driven by the need for a larger atlas, in line with the requirements of large simplicial complexes in the methodologies conjectured by Bobrowski and Skraba for detecting statistically significant loops \cite{BobrowskiKahleSkraba}. The functional connectivity matrices reflect the similarity between these numerous brain regions of interest. The construction of these matrices involved calculating Pearson correlation coefficients between the functional time series of Blood-oxygen-level-dependent imaging (BOLD) activity of different brain regions during the resting-state scans \cite{centeno2022hands}. We focused exclusively on the Young Adult database, considering its relevance, homogeneity compared with ageing databases, and the comprehensive nature of its data. 

 After acquiring the data, we progressed to identifying significant links in the persistent diagrams through several approaches: persistent lengths, persistent ratios, and utilizing the conjectures proposed by Bobrowski and Skraba \cite{BobrowskiKahleSkraba} for the LGumbel distribution, with and without the Bonferroni correction, and a p-value of $0.05$. We calculated the z-scores of their respective distributions for both persistent lengths and ratios. A threshold was then established to differentiate significant cycles from noise. Specifically, a cycle was deemed significant if its z-score surpassed this threshold. This thresholding method to distinguish signal from noise was applied across various threshold levels to test the robustness of this  analysis. The results are summarized in Fig. 1.
 %
 %
 
Intuitively, one could expect a relation between the number of significant loops and the chosen z-score threshold, determining whether a bar in a barcode represents signal or noise. However, our analysis revealed a surprising consistency: although varying z-score points affect the absolute number of significant loops identified, the overall histogram of homologies at the population level remains similar across different z-score thresholds, as can be inferred from our heatmap in Fig. 1. In particular, for a fixed value of $z$, the length and ratio test are always very strongly correlated. 


Given that the distribution of significant loops is strongly correlated for several values of z-scores, including the full set of one-loops (See Fig. 2 (a)), this pattern suggests a scale-invariant, or fractal-like, distribution of significant homologies within the population. 

To investigate this hypothesis further, we calculated the fractal dimension using both i) the box count method \cite{liebovitch1989fast} across the distribution of one-cycles across the population (Fig. 2(b)) and ii) the Hurst exponent \cite{mielniczuk2007estimation} (Fig 2(c)). To compute the Hurst exponent, we initially created a pseudo time series for each individual, quantifying the number of significant loops as a function of various z-score thresholds. Subsequently, we averaged these individual time series across the population to derive a composite time series. The Hurst exponent was calculated using this aggregated data, as illustrated in Fig. 2(c). Notably, similar patterns were observed at the individual level, suggesting a consistent fractal behaviour across the population in the distribution of significant loops, which is robust regarding the threshold used. Our findings give strong evidence of the scale-invariant nature of the distribution of one-cycles at the population level. 

With these two fractal exponents, we have strong evidence of the distribution of the one-homologies at the population level of the Human Connectome Project is fractal. This gives a potential explanation as to why it did not matter which statistical test we used, and they all gave similar results. 

The only value that does not have a very strong correlation  with the rest of the data is the significant cycles with respect to the LGumbel distribution with the Bonferroni correction. In our data, the Bonferonni correction for the LGumbel test lowers the significance level by such an amount, that only a small number of cycles survive. 

Consequently, this paradigmatic exploration on a large dataset challenges the traditional notion of distinguishing between noise and signal in persistent diagrams, given the observed self-similarity in their one-loop distribution.

We further explored the connection between scale invariance and individual behavioural traits. Specifically, we examined the correlation between the number of significant loops in each individual’s brain network and their behavioural brain data, which is available from the Human Connectome Project database \cite{van2012human}. 
Our findings, illustrated in Fig. \ref{fig:brainvolumecorrelation}
, demonstrate a strong correlation between the number of significant loops and brain size. This relationship is evident across different z-score thresholds, but the most pronounced correlation occurs when considering the total number of one-loops without applying any threshold to distinguish between signal and noise. This suggests that all loops, regardless of their persistence length or whether they are classified as signal or noise, contribute meaningfully to this correlation, underscoring the importance of considering the entire set of loops in understanding the relationship between topological features and brain size.

Additionally, as demonstrated in \ref{fig:brainvolumecorrelation}, we observed distinct sex-based differences in loop numbers, with female individuals exhibiting significantly more loops, on average, than male individuals. The robustness of these findings is underscored by the fact that these correlations with brain size and sex differences remain consistent across different threshold levels. 





\begin{figure}[h]\label{fig:brainheatmap}
\centering
    \includegraphics[width=0.8\textwidth]{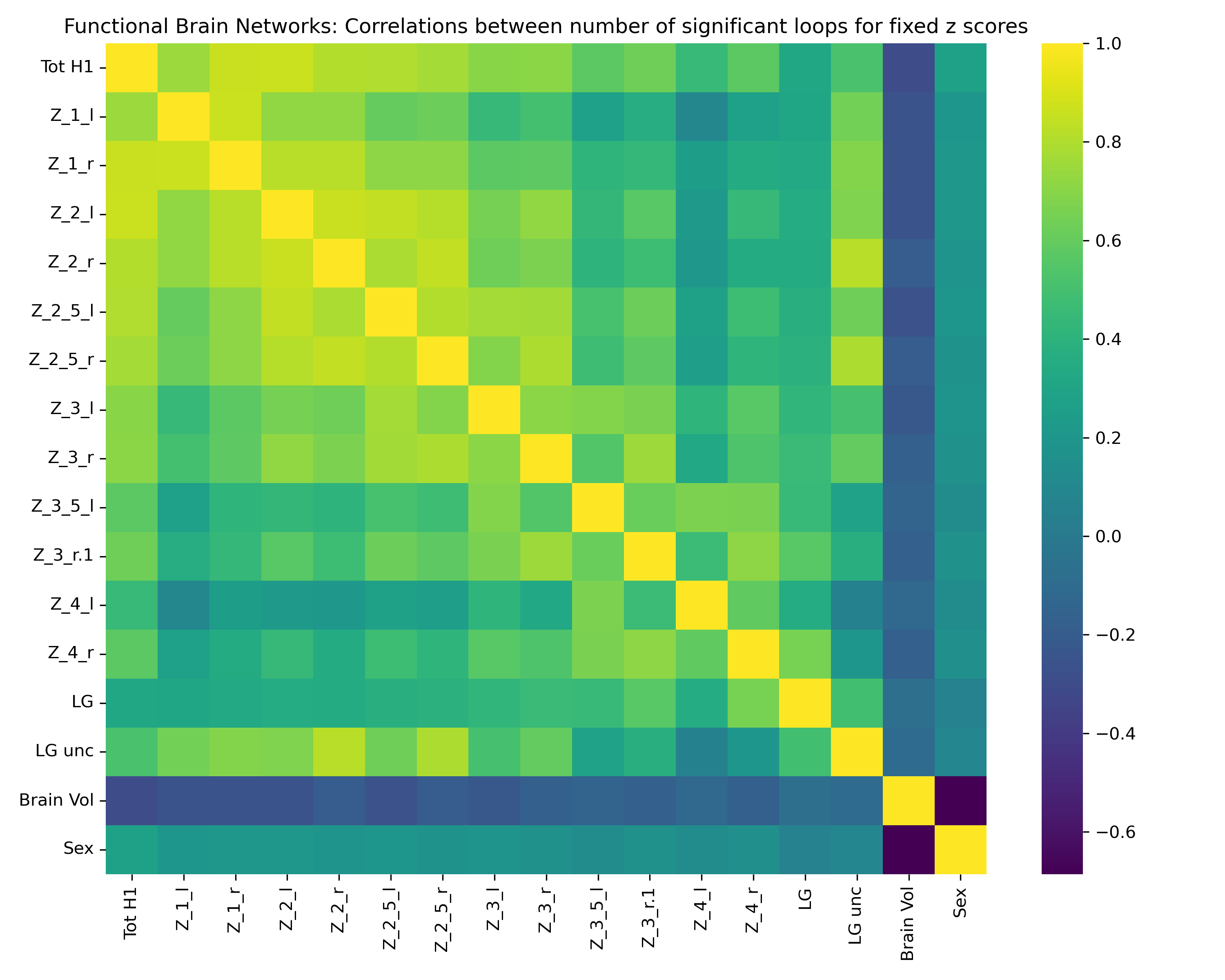}
    \caption{The heatmap corresponds to the correlations between the total number of loops, the number of significant loops using different \(z\)-scores for the persistence length and ratio, and the corrected and uncorrected LGumbel distribution. The values \(z_{n_l}\) (respectively \(z_{n_r}\)) correspond to the total number of loops whose length (respectively ratio) has a \(z\)-score larger than \(n\). The total number of loops is denoted by \(\text{Tot H1}\), \(\text{LG}\) denotes the number of significant loops using the LGumbel distribution with Bonferroni correction, and \(\text{LG unc}\) denotes the number of significant loops using the LGumbel distribution without Bonferroni correction.}
\end{figure}

\begin{figure}[h]
\begin{minipage}[b]{0.31\linewidth}
    \centering
    \includegraphics[width=\linewidth]{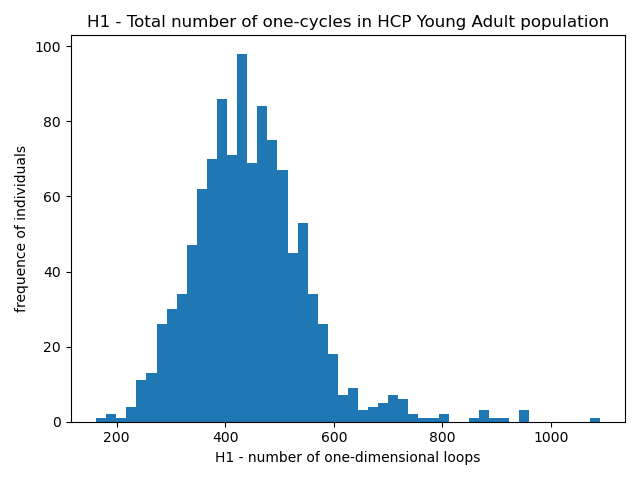}
  \end{minipage}
  \begin{minipage}[b]{0.33\linewidth}
    \centering
    \includegraphics[width=\linewidth]{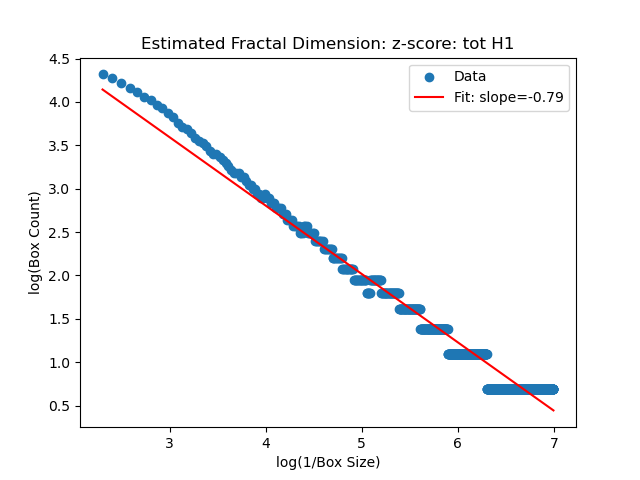}
  \end{minipage}
  \begin{minipage}[b]{0.33\linewidth}
    \centering
    \includegraphics[width=\linewidth]{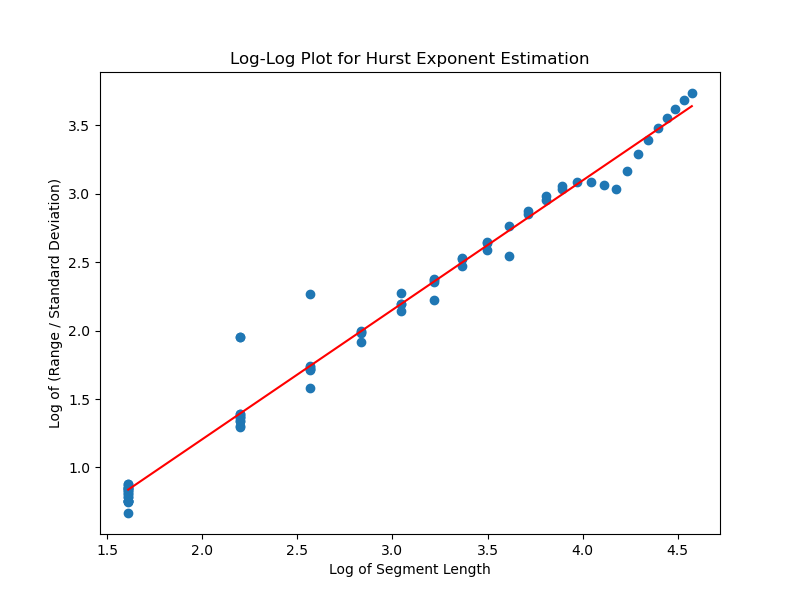}
  \end{minipage}
  \caption{Persistent homology of HCP rs-fMRI data: In (a), we plotted the distribution of the total number of loops; in (b), the estimated fractal dimension using the box counting method for the total number of loops; and in (c), we estimated the Hurst exponent of the same dataset.}
  \label{fig:brainboxcounts}
\end{figure}





\begin{figure}
\centering
\begin{minipage}[b]{0.44\linewidth}
    \centering
    \includegraphics[width=\linewidth]{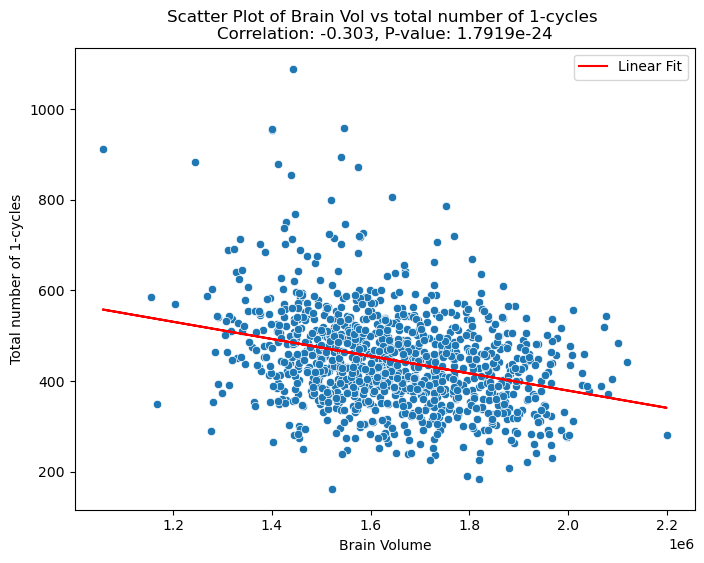}
  \end{minipage}
  \begin{minipage}[b]{0.45\linewidth}
    \centering
    \includegraphics[width=\linewidth]{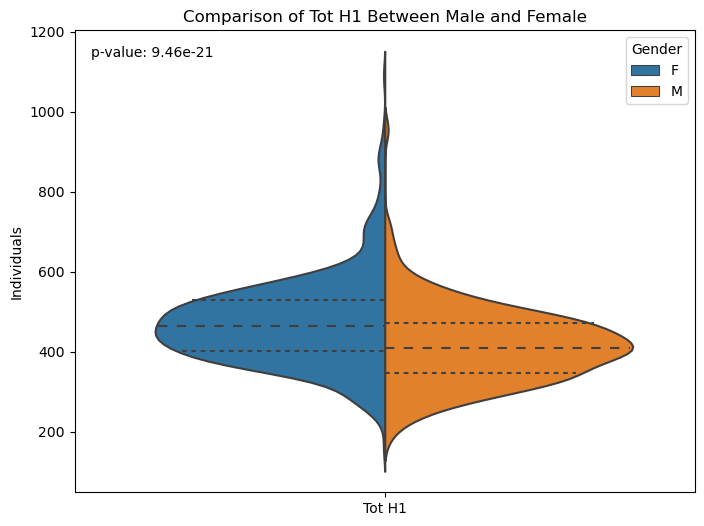}
  \end{minipage}
\label{fig:brainvolumecorrelation}
    \caption{HCP rs-fMRI dataset: (a) Correlation of the total number of one-cycles with brain volume and (b) sex-based differences. Left: A scatter plot illustrating the correlation between the total number of one-cycles in brain networks and individual brain volumes, highlighting the relationship between topological features and brain size. Right: A violin plot comparing the total number of one-cycles between male and female individuals, demonstrating significant sex-based differences in the functional topological structure of brain networks.}
\end{figure}



\section{Daily distribution of 1-cycles is scale invariant and correlates with S\&P 500 Volatility}\label{sec:financialdata}

Topological data analysis, particularly persistent homology, has been increasingly applied in finance across various domains, from understanding market dynamics and identifying financial market crashes to analyzing correlations among assets and optimizing investment strategies \cite{gong2021warning,goel2024sparse,guo2020empirical,rai2024identifying,yen2021understanding}.

We now move to an exploratory analysis of Persistent Homology in financial data sets. 
To this aim, we constructed financial networks based on S\&P 500 data to analyze using TDA. The data was sourced by downloading minute-by-minute trading information for all S\&P 500 tickers listed on Wikipedia, utilizing the Alpha Vantage API \cite{AlphaVantageAPI}. This granular data was then organized daily, forming a time series for each trading day in the S\&P 500.

For every trading day in $2023$, up to $November\mbox{ }23$, as in functional brain networks, we calculated absolute correlation matrices of the returns of each time-series, representing the entire day's trading correlated activity. These matrices, totalling $225$ in number, served as the basic financial networks for our analysis \cite{onnela2004clustering,bardoscia2021physics}. Alongside this, we also obtained the time series of the daily volatility of the S\& P 500. This additional data allowed for a direct comparison between the structural properties of the daily financial networks and the market's daily volatility, providing a comprehensive perspective of the market dynamics.

Using these networks and the volatility data, we applied persistent homology techniques to compute the birth and death times of features within the data and their persistence ratios (death/birth time). 

To standardize and interpret this persistent data, we calculated the $z$-scores for the distribution of both the cycle lengths and their corresponding ratios daily. This step was critical in distinguishing significant topological features from noise. We defined a threshold for the $z$-scores, with values above this threshold indicating significant loops (suggestive of underlying structural patterns) and values below it being suggested as noise. This analysis was conducted across multiple $z$-score thresholds to examine the robustness and consistency of the identified structures.

Upon the publication of this paper, the correlation matrices used in our analyses and the daily volatility data of the S\&P 500 will be made publicly available, fostering further research and verification of our findings.



\begin{figure}
\centering
\label{fig:heatmapstocks}
    \includegraphics[width=0.8\textwidth]{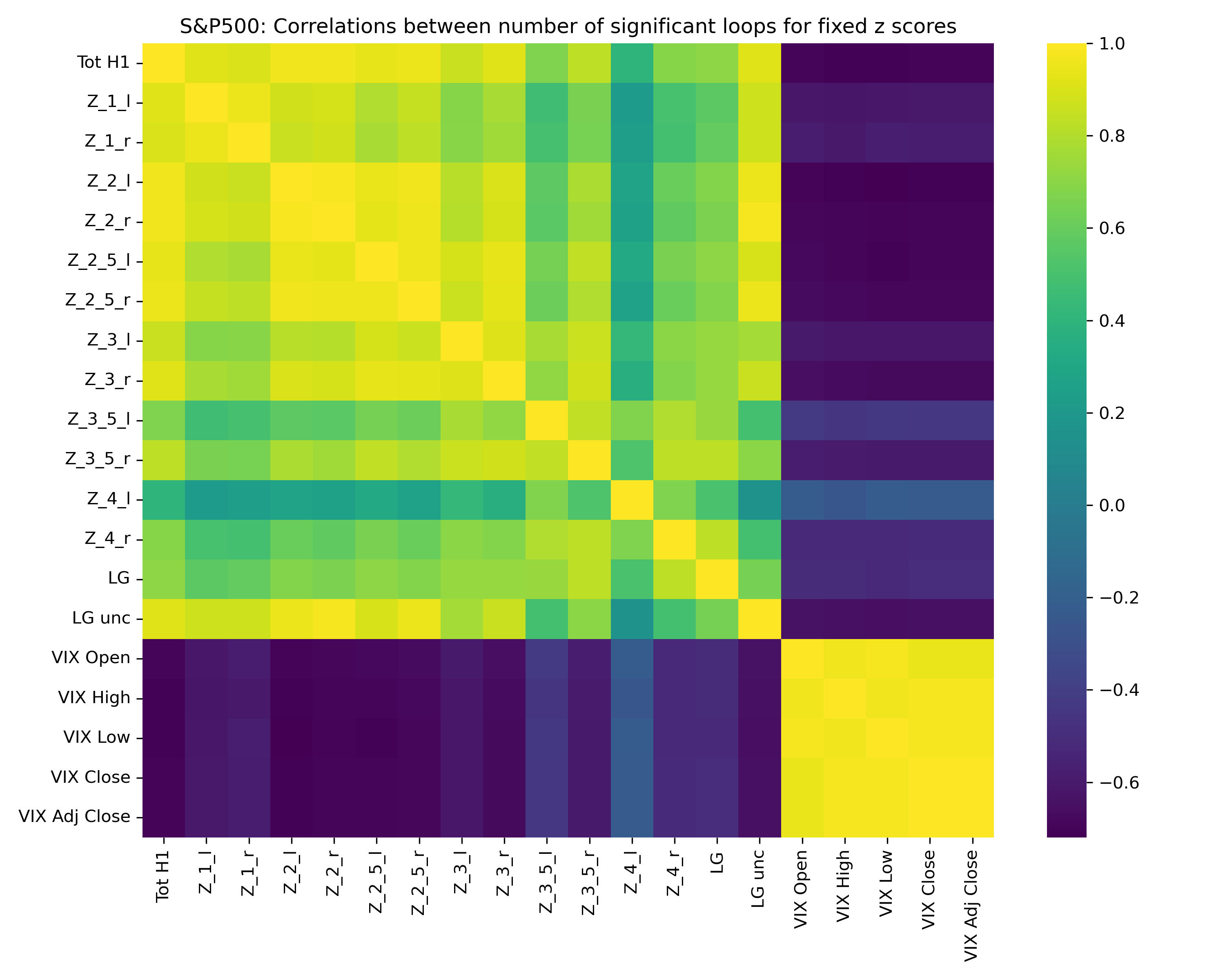}
    \caption{Heatmap of correlation coefficients among different metrics for identifying significant loops and their relation with market volatility for the S\&P500 from all days of 2023 (until 24/11/2023). This heatmap displays the correlation coefficients between the number of significant loops identified by various \(z\)-score thresholds (both for length and ratio distributions) and the Left Gumbel (LG) test (Bonferroni corrected and uncorrected) as proposed by Bobrowski and Skraba \cite{BobrowskiKahleSkraba}. It highlights the strong positive correlations among different loop quantification metrics, regardless of the threshold chosen—a clear signature of scale invariance. Additionally, the heatmap shows a notable negative correlation between the number of significant loops and market volatility, suggesting an inverse relationship between loop prevalence in financial networks and market instability.}
\end{figure}

In Figure \ref{fig:heatmapstocks}, in a similar fashion, as we did for functional brain networks, we see the correlation between the number of loops that pass each of the statistical tests for different values of $z$ and the number of loops significantly differing from the  LGumbel distribution. Similar to the brain data, these strong correlations across all thresholds indicate the possibility of a fractal structure. 

In our financial dataset analysis, we observed a strong correlation in the distribution of significant loops across various z-score thresholds, including when considering the entire set of one-loops (refer to Fig. 3(a)). This consistent pattern indicates a scale-invariant, or fractal-like, distribution of significant homologies in the financial networks.

To further explore this fractal nature, we employed two approaches: i) calculating the fractal dimension using the box count method \cite{liebovitch1989fast} over the one-cycle distribution (Fig. 3(b)), and ii) determining the Hurst exponent \cite{mielniczuk2007estimation} (Fig. 2(c)). For the Hurst exponent calculation, a pseudo time series was generated for each trading day, reflecting the number of significant loops at different z-score thresholds. The average of these daily time series was then used to compute the Hurst exponent, as shown in Fig. 2(c). This approach revealed consistent fractal patterns across individual trading days, reinforcing the evidence of scale invariance in the distribution of one-cycles within the financial market.


\begin{figure}[h]\label{fig:financialboxcounts}
  \begin{minipage}[b]{0.31\linewidth}
    \centering
    \includegraphics[width=\linewidth]{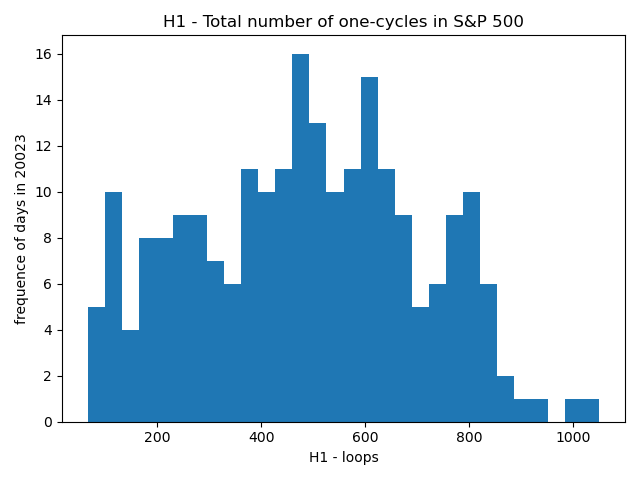}
  \end{minipage}
  \begin{minipage}[b]{0.33\linewidth}
    \centering
    \includegraphics[width=\linewidth]{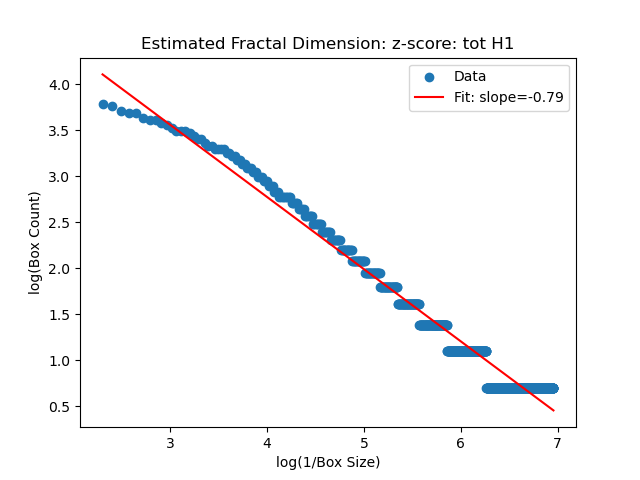}
  \end{minipage}
  \begin{minipage}[b]{0.33\linewidth}
    \centering
    \includegraphics[width=\linewidth]{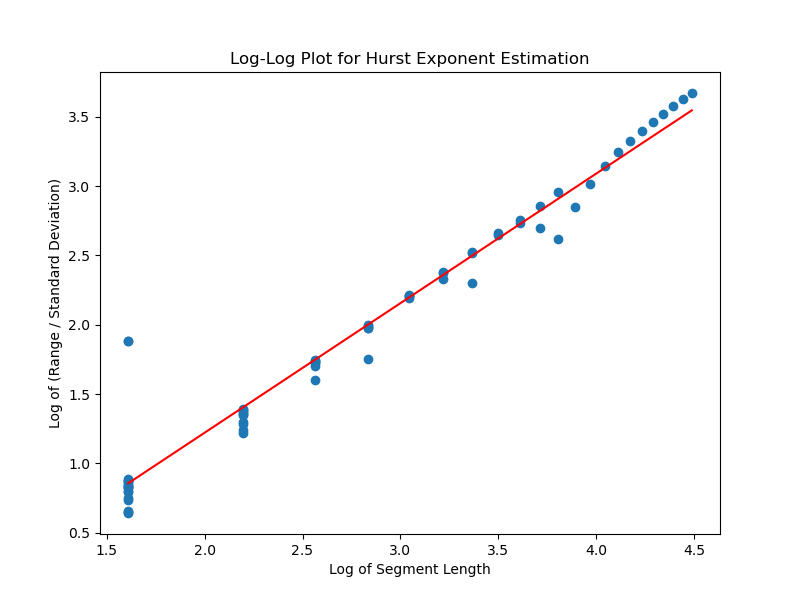}
  \end{minipage}
  \caption{Analysis of one-dimensional cycles and scale invariance in daily financial networks of the S\&P500. (a) Histogram depicting the distribution of the number of H1 loops, illustrating the frequency and range of loop occurrences. (b) Log-log plot of the box count versus box size of H1 loops, indicating scale invariance and fractal behavior in the data. (c) Log-log plot of the Hurst exponent calculated from the average length distribution of H1 loops for all days of 2023, indicating long-range dependence and fractal properties.}
\end{figure}

\begin{figure}[h]\label{fig:volatilitycorrelation}
\centering
\begin{center}
    \includegraphics[width=0.7\textwidth]{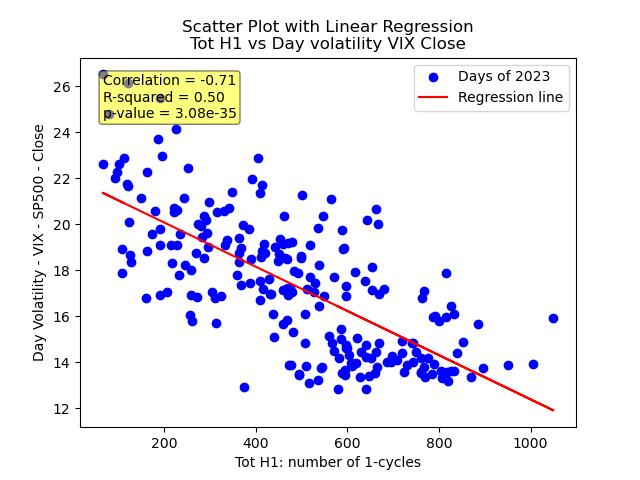}
    \caption{One-cycles correlate with market volatility: This figure presents the relationship between the total number of one-cycles identified in the financial networks of the S\&P 500 on each trading day of 2023 (up to 23/11, totaling 225 trading days) and the corresponding market volatility (VIX) for that day. The plot aims to highlight the extent and nature of the correlation, offering insights into how topological features of financial networks are associated with market fluctuations.}
\end{center}
\end{figure}

\section{Conclusion and discussion}

In this paper, we provided a comparison between two different methods of interpreting persistent homology, which was meant to distinguish signal from noise. In the two data sets we studied, we reached the remarkable conclusion that both persistence length and persistence ratios were strongly correlated for all $z$-values. Thus, in our data sets, all different ways to distinguish signal and noise yielded comparable results. In these data sets, the Bobrowski and Skraba conjecture was therefore of no advantage or disadvantage.

\medskip

A possible explanation for this independence of statistical tests and invariance under different $z$-scores could potentially be because the data is fractal. We expect that, in general, fractal data sets have scale-invariant properties, and therefore that every statistical test will give strongly correlated results. In practice, this would have the consequence that essentially every cycle is part of the signal, or at least that we cannot really distinguish signal and noise in the persistence, as they are entangled. So far, this is just a suspicion based on our empirical evidence, and further theoretical work is necessary to formulate precise conjectures and theorems on the signal and noise in fractally distributed bar codes. We would also like to point out that none of the data sets in \cite{BobrowskiSkraba}, on which Bobrowksi and Skraba tested their conjectures, had any fractal properties. Therefore, they do not serve as counterexamples to the idea that fractal data sets have scale and statistical test invariant properties.

\medskip

The relationship between TDA and fractals is a matter of recent investigations. For example, in \cite{Adamsetal} and \cite{JaquetteSchweinhart}, persistent homology is used to estimate the fractal dimension of point samples and measures. In both these cases, TDA is used to study properties of fractals.  Based on our experiments, we suggest that the converse might also be possible, namely that in certain cases it is possible to use the fractal aspects of barcodes, specifically length and ratio distributions, to find scale invariance in TDA, going beyond the classic task of distinguishing signal from noise.

\medskip

In the comparisons in this paper, we have only focused on the total number of loops or $1$-cycles. There are several further ways to compare the different statistical methods, which we did not do mainly due to the increased computational complexity. First of all, it would be interesting to pick a basis for the persistent homology and compare the explicit cycles that pass each test. So far, we have seen that for each $z$-score, roughly the same number of cycles pass the length and ratio tests. It is unclear whether these are the same cycles or whether different cycles pass the tests. A second way of improving our comparison would be to look at the higher homologies as well. For computational reasons, we have restricted ourselves only to $1$-cycles. However, it would be interesting to see whether the fractal properties of the data sets also appear in the higher homologies.

In summary, our findings suggest that when the distribution of the barcodes within a data set is fractal, then it is unfeasible to distinguish between noise and signal. However, our results represent only initial steps in understanding the complex interplay of signal, noise, and scale-invariant and fractal properties of barcodes in persistent homology. Our study highlights the necessity for further, grounded research to unravel these intricate relationships and validate our preliminary observations. 


\subsection*{Acknowledgements}

The second author is supported by the Dutch Institute for Emergent Phenomena (DIEP) research fellowship. The third author is supported by Dutch Research Organisation (NWO) grant number VI.Veni.202.046 and NWO grant 613.001.651. We are grateful to Katharina Natter for proofreading our manuscript.

\bibliographystyle{plain} 
\bibliography{bibliography}

\end{document}